\documentclass{elsart}

\begin{document}

\begin{frontmatter}

\title{Monotonically equivalent entropies and solution of additivity equation}

\author{Pavel Gorban}

\address{Omsk State University, Omsk, Russia}

\address{ETH Zurich, Department of Materials, Institute of Polymers\\
ETH-Zentrum, CH-8092 Zurich, Switzerland}

\ead{PavelGorban@yandex.ru}

\begin{abstract}
Generalized entropies are studied as Lyapunov functions for the Master equation (Markov chains).
Three basic properties of these Lyapunov functions are taken into consideration: universality
(independence of the kinetic coefficients), trace-form (the form of sum over the states), and
additivity (for composition of independent subsystems). All the entropies, which have all three
properties simultaneously and are defined for positive probabilities, are found. They form a
one-parametric family.

We consider also pairs of entropies $S_{1}$, $S_{2}$, which are connected by the monotonous
transformation $S_{2}=F(S_{1})$ (equivalent entropies). All classes of pairs of universal equivalent
entropies, one of which has a trace-form, and another is additive (these entropies can be different
one from another), were found. These classes consist of two one-parametric families: the family of
entropies, which are equivalent to the additive trace-form entropies, and the family of Renyi-Tsallis
entropies.
\end{abstract}

\end{frontmatter}

\section{\bf Introduction}

The interest to non-classical entropies holds for many decades \cite{Renyi}, \cite{Aczel},
\cite{Tsallis}. Some authors even compiled tables of entropies \cite{EstMor}. This interest is
supported by the successes of ``nonextensive statistical mechanics" in the description of different
phenomena \cite{Abik}, \cite{Beck}.

There are many possible frameworks for the consideration of entropies. One of them is fixed by the
following condition: entropy in isolated systems must grow monotonically with time for every physical
kinetics. In this work we investigate different entropies as Lyapunov functions for Markov chains
\cite{Morimoto}, \cite{Gor-Kar}, \cite{Obhod}.

The basic model we consider here is a finite Markov chain (finiteness and discreteness are by no
means principal restrictions, and are employed only in order to avoid convergence questions).The time
evolution of state probabilities $p_{i}$, where $i$ is the discrete label of the state, is given by
the Master equation
\begin{equation}\label{master}
\dot{p}_{i}=\sum_{j,j\neq i}k_{ij}(\frac{p_{j}}{p_{j}^{*}}-\frac{p_{i}}{p_{i}^{*}})
\end{equation}
where $p_{i}^{*}$ are the equilibrium probabilities and the coefficients must satisfy following
condition
\begin{equation}\label{unitarcon}
\sum_{j,j\neq i}k_{ij}=\sum_{j,j\neq i}k_{ji}
\end{equation}
We consider only the systems which allow positive equilibrium, $p_{i}^{*}>0$ (for infinite systems,
it is often advantageous to use unnormalized $p^{*}$).

A function $H({\mathbf p}, {\mathbf p^{*}})$ (${\mathbf p}=(p_{i})$, ${\mathbf p^{*}}=(p_{i}^{*})$)
is a Lyapunov function for the Markov chain (\ref{master}), if its time derivative in accordance to
(\ref{master}) is non-positively defined ($\frac{d H}{d t}\leq0$).

We will consider three important properties of Lyapunov functions $H(\mathbf{p}, \mathbf{p^{*}})$:

1) {\it Universality}: $H$ is a Lyapunov function for Markov chains (\ref{master}) with a given
equilibrium $\mathbf{p^{*}}$ with every possible values of kinetic coefficients $k_{ij}\geq0$

2) $H$ is a {\it trace-form function}.
\begin{equation}\label{trfrm}
H({\mathbf p },{\mathbf p^{*}})=\sum_{i}f(p_{i},p_{i}^{*})
\end{equation}
where $f$ is a differentiable function of two variables.

3) $H$ is {\it additive} for composition of independent subsystems. It means that if ${\mathbf
p}=p_{ij}=q_{i}r_{j}$ and ${\mathbf p^{*}}=p_{ij}^{*}=q_{i}^{*}r_{j}^{*}$ then $H({\mathbf
p})=H({\mathbf q})+H({\mathbf r})$

Here and further we suppose $0<p_{i},p_{i}^{*},q_{i},q_{i}^{*},r_{i},r_{i}^{*}<1$.

In the next sections we consider the additivity condition as a functional equation and solve it. In
the section 2 we will describe all Lyapunov functions which have all three of these properties
simultaneously and in the section 3 we will find all non-additive trace-form entropies, which become
additive after monotonous transformation of the entropy scale.

\section{\bf Non-classic additive entropies}

The following theorem describes all Lyapunov functions for Markov chains, which have all three
properties 1) - 3) simultaneously.

Let $f(p,p^{*})$ be a twice differentiable function of two variables.

{\it Theorem 1.} If a function $H({\mathbf p},{\mathbf p^{*}})$ has all the properties 1)-3)
simultaneously, then
\begin{equation}\label{th1eq1}
f(p,p^{*})=p_{i}^{*}h(\frac{p}{p^{*}}),\mbox{ }\mbox{ } H({\mathbf p},{\mathbf p^{*}})=
\sum_{i}p_{i}^{*}h(\frac{p_{i}}{p_{i}^{*}})
\end{equation}
where
\begin{equation}\label{th1eq2}
h(x)=C_{1}\ln x+C_{2}x\ln x,\mbox{ }C_{1}\leq0,\mbox{ }C_{2}\geq0
\end{equation}

{\it Lemma 1}. If a Lyapunov function H(p) for the Markov chain is of the trace-form
($H(p)=\sum_{i}f(p_{i},p_{i}^{*})$) and is universal, then $f(p,p^{*})=p^{*}h(\frac{p}{p^{*}})$,
where $h(x)$ is a convex function of one variable.

{\it Proof}. Let's consider a Markov chain with two states. For such a chain
$\dot{p}_{1}=q_{12}p_{2}^{*}(\frac{p_{2}}{p_{2}^{*}}-\frac{p_{1}}{p_{1}^{*}})=-q_{21}p_{1}^{*}(\frac{p_{1}}{p_{1}^{*}}-\frac{p_{2}}{p_{2}^{*}})=-\dot{p}_{2}$,
(here $q_{ij}p_{j}^{*}=k_{ij}$) so as we have $\dot{H}\leq0$ for $H$ to be a Lyapunov function, we
get an equation $(\frac{\partial f(p_{2},p_{2}^{*})}{\partial p_{2}}-\frac{\partial
f(p_{1},p_{1}^{*})}{\partial p_{1}}))(\frac{p_{1}}{p_{1}^{*}}-\frac{p_{2}}{p_{2}^{*}})\leq0$, from
which we can see that $\frac{\partial f(p,p^{*})}{\partial p}$ is a monotonous function of
$\frac{p}{p^{*}}$, so $f(p,p^{*})=p^{*}h(\frac{p}{p^{*}})$, where $h$ is a convex function of one
variable.

Lemma 1 is proven.

{\it Proof of the theorem 1}. Let $H$ have the form (\ref{th1eq1}) and $h$ to be twice differentiable
in the interval $(0,+\infty)$. Then the additivity equation
\begin{equation}\label{addeq}
H({\mathbf p})-H({\mathbf q})-H({\mathbf r})=0,
\end{equation}
holds. Here (in (\ref{addeq}))
\begin{eqnarray*}
&&q_{n}=1-\sum_{i=1}^{n-1}q_{i}, r_{m}=1-\sum_{j=1}^{m-1}r_{j},\\ &&H({\mathbf
p})=\sum_{i,j}q_{i}^{*}r_{j}^{*}h(\frac{q_{i}r_{j}}{q_{i}^{*}r_{j}^{*}}), H({\mathbf
q})=\sum_{i}q_{i}^{*}h(\frac{q_{i}}{q_{i}^{*}}),\\&& H({\mathbf
r})=\sum_{j}r_{j}^{*}h(\frac{r_{j}}{r_{j}^{*}}), {\mathbf p}=p_{ij}=q_{i}r_{j}.
\end{eqnarray*}
Let's take the derivatives of this equation first on $q_{1}$ and then on $r_{1}$. Then we get the
equation ($g(x)=h'(x)$)
\begin{eqnarray*}
&g(\frac{q_{1}r_{1}}{q_{1}^{*}r_{1}^{*}})-g(\frac{q_{n}r_{1}}{q_{n}^{*}r_{1}^{*}})-
g(\frac{q_{1}r_{m}}{q_{1}^{*}r_{m}^{*}})+g(\frac{q_{n}r_{m}}{q_{n}^{*}r_{m}^{*}})+\\
&+\frac{q_{1}r_{1}}{q_{1}^{*}r_{1}^{*}}g'(\frac{q_{1}r_{1}}{q_{1}^{*}r_{1}^{*}})-
\frac{q_{n}r_{1}}{q_{n}^{*}r_{1}^{*}}g'(\frac{q_{n}r_{1}}{q_{n}^{*}r_{1}^{*}})-
\frac{q_{1}r_{m}}{q_{1}^{*}r_{m}^{*}}g'(\frac{q_{1}r_{m}}{q_{1}^{*}r_{m}^{*}})+
\frac{q_{n}r_{m}}{q_{n}^{*}r_{m}^{*}}g'(\frac{q_{n}r_{m}}{q_{n}^{*}r_{m}^{*}})=0
\end{eqnarray*}
Lets denote $x=\frac{q_{1}r_{1}}{q_{1}^{*}r_{1}^{*}}$, $y=\frac{q_{n}r_{1}}{q_{n}^{*}r_{1}^{*}}$,
$z=\frac{q_{1}r_{m}}{q_{1}^{*}r_{m}^{*}}$, and $\psi(x)=g(x)+xg'(x)$. It's obvious that if $n$ and
$m$ are more than 2, then $x$, $y$ and $z$ are independent and can take any positive values. So, we
get the functional equation:
\begin{equation}
\psi(\frac{yz}{x})=\psi(y)+\psi(z)-\psi(x)
\end{equation}
Let's denote $C_{2}=-\psi(1)$ and $\psi_{1}(\alpha)=\psi(\alpha)-\psi(1)$ and take $x=1$. We get then
\begin{equation}
\psi_{1}(yz)=\psi_{1}(y)+\psi_{1}(z)
\end{equation}
the Cauchy functional equation. The solution of this equation in the class of measurable functions is
$\psi_{1}(\alpha)=C_{1}\ln\alpha$, where $C_{1}$ is constant. So we get $\psi(x)=C_{1}\ln x+C_{2}$
and $g(x)+xg'(x)=C_{1}\ln x+C_{2}$. The solution is $g(x)=\frac{C_{3}}{x}+C_{1}\ln x+C_{2}-C_{1}$;
$h(x)=\int(\frac{C_{3}}{x}+C_{1}\ln x+C_{2}-C_{1})dx=C_{3}\ln x+C_{1}x\ln x+(C_{2}-2C_{1})x+C_{4}$,
or, renaming constants, $h(x)=C_{1}\ln x+C_{2}x\ln x+C_{3}x+C_{4}$. In the expression for $h(x)$
there are two parasite constants $C_{3}$ and $C_{4}$ which occurs because the initial equation was
differentiated twice. So, $C_{3}=0$, $C_{4}=0$ and $h(x)=C_{1}\ln x+C_{2}x\ln x$. Because $h$ is
convex, we have $C_{1}\leq0$ and $C_{2}\geq0$.

Theorem 1 is proven.

In the case when $C_{1}=0$, the function $H(x)$ is continuous in the interval $[0,+\infty)$. When
$C_{1}\neq0$ the function $h(x)$ is  defined only in the open interval $(0,+\infty)$. Because of this
reason one can easily not to consider the ``Burg part" of the entropy $H(p)=\sum_{i}p_{i}(C_{1}\ln
\frac{p_{i}}{p_{i}^{*}}+C_{2}\frac{p_{i}}{p_{i}^{*}}\ln \frac{p_{i}}{p_{i}^{*}})$ under an assumption
that $H(p)$ is continuous if some of $p_{i}$ are equal to zero \cite{Aczel}.

There is one more entropy, called Hartley entropy, which can be written as $S({\mathbf
p})=\sum_{i}sign(p_{i})$, where $sign(x)=1$ if $x>0$ and $sign(x)=0$ if $x\leq0$. It is additive and
it is of trace-form, but we have not found it because it is constant when $p_{i}>0$.

Family of the Lyapunov functions (\ref{th1eq1}), (\ref{th1eq2}) was introduced for the first time in
the book of A.Gorban \cite{Obhod}. The applicability of this entropy for the universal description of
some systems out of thermodynamic limit was considered in the work \cite{Gor-Kar}. The part $C_{1}\ln
x$ corresponds to the Burg entropy \cite{Burg}, the part $C_{2}x\ln x$ corresponds to the classical
Boltzmann-Gibbs-Shannon entropy.

\section{\bf Trace-form entropies, which become additive after the monotonous transformation}

There is a further class of very useful entropies $S({\mathbf p})$, which are of the trace-form, but
not additive, and after some monotonous transformation $S_{1}({\mathbf p})=F(S({\mathbf p}))$ become
additive, but loose their trace-form (such entropies $S({\mathbf p})$ and $S_{1}({\mathbf p})$ are in
fact equivalent (from thermodynamic point of view), because the sets of isoentropic levels coincide).
One well known example of such pair of entropies is ($q\neq0$)
\begin{eqnarray*}
S({\mathbf p})=S_{q}^{(T)}({\mathbf p})=\frac{1}{1-q}\left(\sum_{i=1}^{n}p_{i}^{q}-1\right)& \mbox{ -
Tsallis entropy and}\\ S_{1}({\mathbf p})=S_{q}^{(R)}({\mathbf
p})=\frac{1}{1-q}\ln\left(\sum_{i=1}^{n}p_{i}^{q}\right)& \mbox{ - Renyi entropy}&\\ \mbox{Here
}S_{1}({\mathbf p})=\frac{1}{1-q}\ln((1-q)S({\mathbf p})+1)&
\end{eqnarray*}
So, there is a problem of finding all the entropies with such properties. We will give an answer
here.

The relation between Renyi entropy and Tsallis entropy was discussed many times. For example, S.Abe
wrote ``even starting with a nonextensive entropy, the thermodynamic entropy appearing at the
macroscopic level has to be extensive, in accordance with Carathedory's theorem. This fact can be
observed as transmutation from nonextensive Tsallis theory to extensive Renyi-entropy-based theory"
\cite{Abik2}.

If two entropies $S_{1}({\mathbf p})$ and $S_{2}({\mathbf p})$ are connected by the monotonous
transformation
\begin{equation}\label{mon}
S_{1}({\mathbf p})=F(S_{2}({\mathbf p})), S_{2}({\mathbf p})=F^{-1}(S_{1}({\mathbf p}))
\end{equation}
then solutions of two following conditional equilibrium problems coincide
\begin{eqnarray}\label{QE}
\nonumber S_{1}({\mathbf p})\rightarrow max,\mbox{ }m({\mathbf p})={\mathbf M}\\ \\ S_{2}({\mathbf
p})\rightarrow max,\mbox{ }m({\mathbf p})={\mathbf M}\nonumber
\end{eqnarray}
Here ${\mathbf M}$ are ``macroscopic variables". Let's denote the solution of (\ref{QE}) as ${\mathbf
p_{M}}$ and $S_{1,2}({\mathbf M})=S_{1,2}({\mathbf p_{M}})$. The connection (\ref{mon}) holds for
this ``macroscopic" entropies (it's obvious):
\begin{equation}\label{monqe}
S_{2}({\mathbf M})=F(S_{1}({\mathbf p_{M}})), S_{1}({\mathbf M})=F^{-1}(S_{2}({\mathbf p_{M}}))
\end{equation}
If entropies are connected by the monotonous transformation, then they are equivalent for solution of
the conditional equilibrium problems (\ref{QE}). From this point of view nonextensive Tsallis
thermodynamics is equivalent to the extensive Renyi thermodynamics (the discussion of sources of
``extensivity of nonextensive thermodynamics" can be found also in the recent paper \cite{ViPl}). If
one is interested not only in quasiequilibrium distributions (\ref{QE}), but also in numerical means
of the entropies, then the differences between equivalent (\ref{mon}) entropies can be essential. For
example, difference between Renyi and Tsallis entropies are displayed when one considers the
stability of entropies in the thermodynamic limit \cite{Abik3}. Nevertheless, classes of
monotonically equivalent entropies (\ref{mon}) in many aspects  act as one single entropy.

We will look for all the classes of monotonically equivalent entropies, in which there is at least
one trace-form entropy and one additive entropy. As in the previous part of the work, we consider
non-increasing Lyapunov functions for Markov chains (minus-entropies).

The problem is to find all such universal and trace-form Lyapunov functions $H$ for Markov chains,
that there exists a monotonous function F, such that $F(H({\mathbf p}))=F(H({\mathbf
q}))+F(H({\mathbf r}))$ if ${\mathbf p}=p_{ij}=q_{i}r_{j}$.

With the Lemma 1 we get that $H({\mathbf
p})=\sum_{i,j}q_{i}^{*}r_{j}^{*}h\left(\frac{q_{i}r_{j}}{q_{i}^{*}r_{j}^{*}}\right)$, $H({\mathbf
q})=\sum_{i}q_{i}^{*}h\left(\frac{q_{i}}{q_{i}^{*}}\right)$, $H({\mathbf
r})=\sum_{j}r_{j}^{*}h\left(\frac{r_{j}}{r_{j}^{*}}\right)$. Let $F(x)$ and $h(x)$ be differentiable
as many times as needed. Differentiating the equality $F(H({\mathbf p}))=F(H({\mathbf
q}))+F(H({\mathbf r}))$ on $r_{1}$ and $q_{1}$ taking into account that
$q_{n}=1-\sum_{i=1}^{n-1}q_{i}$ and $r_{m}=1-\sum_{j=1}^{m-1}r_{j}$ we get that $F'(H({\mathbf
p}))H''_{q_{1}r_{1}}({\mathbf p})=-F''(H({\mathbf p}))H'_{q_{1}}({\mathbf p})H'_{r_{1}}({\mathbf
p})$, or, if $-\frac{F'(H({\mathbf p}))}{F''(H({\mathbf p}))}=G(H({\mathbf p}))$ then
\begin{equation}
G(H({\mathbf p}))=\frac{H'_{q_{1}}({\mathbf p})H'_{r_{1}}({\mathbf p})}{H''_{q_{1}r_{1}}({\mathbf
p})}
\end{equation}
It is possible if and only if every linear differential operator of the first order, which annulates
$H({\mathbf p})$ and $\sum p_{i}$ annulate also
\begin{equation}\label{Geqn}
\frac{H'_{q_{1}}({\mathbf p})H'_{r_{1}}({\mathbf p})}{H''_{q_{1}r_{1}}({\mathbf p})}
\end{equation}
and it means that every differential operator which has the form
\begin{equation}\label{Dfrm}
D=\left(\frac{\partial H({\mathbf p})}{\partial q_{\gamma}}-\frac{\partial H({\mathbf p})}{\partial
q_{\alpha}}\right)\frac{\partial}{\partial q_{\beta}}+\left(\frac{\partial H({\mathbf p})}{\partial
q_{\beta}}-\frac{\partial H({\mathbf p})}{\partial q_{\gamma}}\right)\frac{\partial}{\partial
q_{\alpha}}+\left(\frac{\partial H({\mathbf p})}{\partial q_{\alpha}}-\frac{\partial H({\mathbf
p})}{\partial q_{\beta}}\right)\frac{\partial}{\partial q_{\gamma}}
\end{equation}
annulates (\ref{Geqn}). For $\beta=2, \alpha=3, \gamma=4$  we get the following equation
\begin{eqnarray}\label{Diffur}
\nonumber F_{1}({\mathbf q},{\mathbf
r})\left[h'\left(\frac{q_{2}r_{1}}{q_{2}^{*}r_{1}^{*}}\right)-h'\left(\frac{q_{2}r_{m}}{q_{2}^{*}r_{m}^{*}}\right)+
\frac{q_{2}r_{1}}{q_{2}^{*}r_{1}^{*}}h''\left(\frac{q_{2}r_{1}}{q_{2}^{*}r_{1}^{*}}\right)-
\frac{q_{2}r_{m}}{q_{2}^{*}r_{m}^{*}}h''\left(\frac{q_{2}r_{m}}{q_{2}^{*}r_{m}^{*}}\right)\right]+\\
\nonumber F_{2}({\mathbf q},{\mathbf
r})\left[h'\left(\frac{q_{3}r_{1}}{q_{3}^{*}r_{1}^{*}}\right)-h'\left(\frac{q_{3}r_{m}}{q_{3}^{*}r_{m}^{*}}\right)+
\frac{q_{3}r_{1}}{q_{3}^{*}r_{1}^{*}}h''\left(\frac{q_{3}r_{1}}{q_{3}^{*}r_{1}^{*}}\right)-
\frac{q_{3}r_{m}}{q_{3}^{*}r_{m}^{*}}h''\left(\frac{q_{3}r_{m}}{q_{3}^{*}r_{m}^{*}}\right)\right]+\\
\nonumber F_{3}({\mathbf q},{\mathbf
r})\left[h'\left(\frac{q_{4}r_{1}}{q_{4}^{*}r_{1}^{*}}\right)-h'\left(\frac{q_{4}r_{m}}{q_{4}^{*}r_{m}^{*}}\right)+
\frac{q_{4}r_{1}}{q_{4}^{*}r_{1}^{*}}h''\left(\frac{q_{4}r_{1}}{q_{4}^{*}r_{1}^{*}}\right)-
\frac{q_{4}r_{m}}{q_{4}^{*}r_{m}^{*}}h''\left(\frac{q_{4}r_{m}}{q_{4}^{*}r_{m}^{*}}\right)\right]=0\\
\end{eqnarray}

where

\begin{eqnarray}
\nonumber F_{1}({\mathbf q},{\mathbf
r})=\sum_{j}r_{j}\left[h'\left(\frac{q_{4}r_{j}}{q_{4}^{*}r_{j}^{*}}\right)-h'\left(\frac{q_{3}r_{j}}{q_{3}^{*}r_{j}^{*}}\right)\right]\mbox{;}\\
\nonumber F_{2}({\mathbf q},{\mathbf
r})=\sum_{j}r_{j}\left[h'\left(\frac{q_{2}r_{j}}{q_{2}^{*}r_{j}^{*}}\right)-h'\left(\frac{q_{4}r_{j}}{q_{4}^{*}r_{j}^{*}}\right)\right]\mbox{;}\\
\nonumber F_{3}({\mathbf q},{\mathbf
r})=\sum_{j}r_{j}\left[h'\left(\frac{q_{3}r_{j}}{q_{3}^{*}r_{j}^{*}}\right)-h'\left(\frac{q_{2}r_{j}}{q_{2}^{*}r_{j}^{*}}\right)\right]\mbox{.}\\
\nonumber
\end{eqnarray}

If we apply the differential operator $\frac{\partial}{\partial r_{2}}-\frac{\partial}{\partial
r_{3}}$, which annulate our conservation law $\sum_{j}r_{j}=1$, to the left part of (\ref{Diffur}),
and denote $f(x)=xh''(x)+h'(x)$, $x_{1}=\frac{q_{2}}{q_{2}^{*}}$, $x_{2}=\frac{q_{3}}{q_{3}^{*}}$,
$x_{3}=\frac{q_{4}}{q_{4}^{*}}$, $y_{1}=\frac{r_{1}}{r_{1}^{*}}$, $y_{2}=\frac{r_{m}}{r_{m}^{*}}$,
$y_{3}=\frac{r_{2}}{r_{2}^{*}}$, $y_{4}=\frac{r_{3}}{r_{3}^{*}}$, we get the equation
\begin{eqnarray}\label{FirstFeq}
\nonumber (f(x_{3}y_{3})-f(x_{2}y_{3})-f(x_{3}y_{4})+f(x_{2}y_{4}))(f(x_{1}y_{1})-f(x_{1}y_{2}))+\\
(f(x_{1}y_{3})-f(x_{3}y_{3})-f(x_{1}y_{4})+f(x_{3}y_{4}))(f(x_{2}y_{1})-f(x_{2}y_{2}))+\\ \nonumber
(f(x_{2}y_{3})-f(x_{1}y_{3})-f(x_{2}y_{4})+f(x_{1}y_{4}))(f(x_{3}y_{1})-f(x_{3}y_{2}))=0
\end{eqnarray}
or, after differentiation on $y_{1}$ and $y_{3}$ and denotation $g(x)=f'(x)$
\begin{eqnarray}\label{SecFeq}
& &x_{1}g(x_{1}y_{1})(x_{3}g(x_{3}y_{3})-x_{2}g(x_{2}y_{3}))+x_{2}g(x_{2}y_{1})(x_{1}g(x_{1}y_{3})-\\
\nonumber & &-x_{3}g(x_{3}y_{3}))+ x_{3}g(x_{3}y_{1})(x_{2}g(x_{2}y_{3})-x_{1}g(x_{1}y_{3}))=0
\end{eqnarray}
If $y_{3}=1$, $y_{1}\neq0$, $\varphi(x)=xg(x)$, we get after multiplication (\ref{SecFeq}) on $y_{1}$
\begin{equation}\label{TrdFeq}
\varphi(x_{1}y_{1})(\varphi(x_{3})-\varphi(x_{2}))+\varphi(x_{2}y_{1})(\varphi(x_{1})-\varphi(x_{3}))+\varphi(x_{3}y_{1})(\varphi(x_{2})-\varphi(x_{1}))=0
\end{equation}
It implies that for every three positive numbers $\alpha$, $\beta$, $\gamma$ the functions
$\varphi(\alpha x)$, $\varphi(\beta x)$, $\varphi(\gamma x)$ are lineary dependent, and for
$\varphi(x)$ the differential equation
\begin{eqnarray*}
ax^{2}\varphi''(x)+bx\varphi'(x)+c\varphi(x)=0
\end{eqnarray*}
holds. This differential equation has solutions of two kinds:

1) $\varphi(x)=C_{1}x^{k_{1}}+C_{2}x^{k_{2}}$, $k_{1}\neq k_{2}$, $k_{1}$ and $k_{2}$ are real or
complex-conjugate numbers.

2) $\varphi(x)=C_{1}x^{k}+C_{2}x^{k}\ln x$.

Let's check, which of these solutions satisfy the equation (\ref{TrdFeq}).

1) $\varphi(x)=C_{1}x^{k_{1}}+C_{2}x^{k_{2}}$. After substitution of this into (\ref{TrdFeq}) and
calculations we get
$C_{1}C_{2}(y_{1}^{k_{1}}-y_{1}^{k_{2}})(x_{1}^{k_{1}}x_{3}^{k_{2}}-x_{1}^{k_{1}}x_{2}^{k_{2}}+x_{1}^{k_{2}}x_{2}^{k_{1}}-x_{2}^{k_{1}}x_{3}^{k_{2}}+x_{2}^{k_{2}}x_{3}^{k_{1}}-x_{1}^{k_{2}}x_{3}^{k_{1}})=0$.
It means that $C_{1}=0$, or $C_{2}=0$, or $k_{1}=0$, or $k_{2}=0$ and the solution of this kind can
have only the form $\varphi(x)=C_{1}x^{k}+C_{2}$.

2) $\varphi(x)=C_{1}x^{k}+C_{2}x^{k}\ln x$. After substitution of this into (\ref{TrdFeq}) and some
calculations if $y_{1}\neq0$ we get $C_{2}^{2}((x_{1}^{k}-x_{2}^{k})x_{3}^{k}\ln
x_{3}+(x_{3}^{k}-x_{1}^{k})x_{2}^{k}\ln x_{2}+(x_{2}^{k}-x_{3}^{k})x_{1}^{k}\ln x_{1})=0$. It mean
that or $C_{2}=0$ and the solution is $\varphi(x)=C_{1}x^{k}$, or $k=0$ and the solution is
$\varphi(x)=C_{1}+C_{2}\ln x$.

So, the equation (\ref{TrdFeq}) has two kinds of solutions:

1)$\varphi(x)=C_{1}x^{k}+C_{2}$

2)$\varphi(x)=C_{1}+C_{2}\ln x$.

Lets solve the equation $f(x)=xh''(x)+h'(x)$ for each of these two cases.

1)$\varphi(x)=C_{1}x^{k}+C_{2}$; $g(x)=C_{1}x^{k-1}+\frac{C_{2}}{x}$ there are two possibilities.

1.1)$k=0$. Then $g(x)=\frac{C}{x}$, $f(x)=C\ln x+C_{1}$, $h(x)=C_{1}x\ln x+C_{2}\ln x+C_{3}x+C_{4}$.

1.2)$k\neq0$. Then $f(x)=Cx^{k}+C_{1}\ln x+C_{2}$, and here are also two possibilities.

1.2.1)$k=-1$. Then $h(x)=C_{1}\ln^{2}x+C_{2}x\ln x+C_{3}\ln x+C_{4}x+C_{5}$.

1.2.2)$k\neq-1$. Then $h(x)=C_{1}x^{k+1}+C_{2}x\ln x+C_{3}\ln x+C_{4}x+C_{5}$

2)$\varphi(x)=C_{1}+C_{2}\ln x$; $g(x)=C_{1}\frac{\ln x}{x}+\frac{C_{2}}{x}$;
$f(x)=C_{1}\ln^{2}x+C_{2}\ln x+C_{3}$; $h(x)=C_{1}x\ln^{2}x+C_{2}x\ln x+C_{3}\ln x+C_{4}x+C_{5}$

(We have renamed constants during the calculations).

For the next step let's see, which of these solutions remains good for equation (\ref{Diffur}). The
result is that there are just two families of functions $h(x)$ such, that equation (\ref{Diffur})
holds:

1)$h(x)=Cx^{k}+C_{1}x+C_{2}$, $k\neq0$, $k\neq1$, and

2)$h(x)=C_{1}x\ln x+C_{2}\ln x+C_{3}x+C_{4}$

From these equalities it is easy to see, that all universal Lyapunov functions $H$ for Markov chains,
for which a monotonous function $F$ exists, such that $F(H({\mathbf p}))=F(H({\mathbf
q}))+F(H({\mathbf r}))$ if ${\mathbf p}=p_{ij}=q_{i}r_{j}$, have the form
\begin{equation}\label{ReTs}
H({\mathbf p})=C\sum_{i}p_{i}^{*}(\frac{p_{i}}{p_{i}^{*}})^{k}+C_{1}
\end{equation}
where $k\neq0$ and $k\neq1$, $C\geq0$, or of the form
\begin{equation}\label{AdTr}
H({\mathbf
p})=\sum_{i}p_{i}^{*}(C_{1}\frac{p_{i}}{p_{i}^{*}}\ln\frac{p_{i}}{p_{i}^{*}}+C_{2}\ln\frac{p_{i}}{p_{i}^{*}})+C_{3}
\end{equation}
where $C_{1}\leq0$ and $C_{2}\geq0$.

So, there are two families of such entropies:

1) Classes of equivalent entropies of Renyi - Tsallis (\ref{ReTs});

2) Classes of entropies, which are monotonically equivalent to additive trace-form entropies
(\ref{AdTr}).

\section{\bf One remark about dimension}

Formally in our problems we considered Lyapunov functions in the form $H({\mathbf p},{\mathbf
p^{*}})$ for Markov chains with every finite number of states. But in the proofs it's enough to use
only Markov chains with a small number of states:

1) In the Lemma 1 (transition from the function of two variables $f(p,p^{*})$ to
$p^{*}h(\frac{p}{p^{*}})$) it's enough to consider Markov chains with two states.

2) In the Theorem 1 (the description of all additive trace-form Lyapunov functions for Markov chains)
it is enough to consider subsystems with 3 states and unification of two independent systems, each of
which has three states.

3) In the last result (the characterization of all classes of monotonically equivalent entropies, in
which there is one trace-form entropy and one additive entropy) it is enough to consider systems with
6 states and their unification (probably, it is enough to consider systems with 4 states, but it is
not proven).

\section{\bf Concluding remarks}

Axiomatic characterization of entropy by it's properties is a very old and well investigated topic
\cite{AczDar}. The development of nonextensive thermodynamics forces us to revisit old questions
\cite{Abik4}, \cite{Kaniadakis}. In particular, it is the question about the frames of investigation
and using of the entropies.

We consider entropy as a Lyapunov function for Markov chains and as a measure of deviation from
equilibrium. In the class of such functions we have found all the entropies, which have the following
three properties simultaneously: universality, trace-form and additivity. This is a family of
additive entropies \cite{Obhod}, \cite{Gor-Kar}, \cite{GorKarOet}. Weaker restriction for the entropy
is that in one scale the entropy has the trace-form, and after the monotonous transformation of the
scale it becomes additive (but can loose the trace-form). There is one more class of monotonically
equivalent entropies, which satisfies these restrictions - the class of Renyi-Tsallis entropies.

{\bf Acknowledgements.} The author would like to thank Dr. I.Karlin for his interest and advises,
Prof. Dr. A.Gorban for the productive discussions of the topic, Prof. Dr. H.C.Oettinger for his
critical view at the paper, S.Ansumali and Dr.A.Zinovyev for their helpful advises on the manuscript.

\end{document}